\documentclass[twocolumn,aps,prl,showpacs,amssymb]{revtex4}

\usepackage{graphicx}
\usepackage{dcolumn}
\usepackage{bm}

\newcommand{\ep}{\varepsilon}
\def\bsigma{\mbox{\boldmath$\sigma$}}

\begin{document}

\preprint{}

\title{Role of surface anisotropy for magnetic impurities in electron
  dephasing and energy relaxation and their size effect}

\author{O. \'Ujs\'aghy$^a$}
 \author{A. Jakov\'ac$^{a}$}
\author{A. Zawadowski$^{a,b}$}
\affiliation{$^{a}$Budapest University of Technology and Economics,
Institute of Physics and Research group Theory of Condensed Matter of 
Hungarian Academy of Sciences
H-1521 Budapest, Hungary}
\affiliation{$^{b}$Research Institute for Solid State Physics, POB 49, H-1525
Budapest, Hungary}

\date{\today}

\begin{abstract}
  Recently the electron dephasing and energy relaxation due to different
  magnetic impurities have been extensively investigated experimentally in thin
  wires and in this Letter these quantities are theoretically studied.  It was
  shown earlier that a magnetic impurity in a metallic host with strong
  spin-orbit interaction experiences a surface anisotropy of the form $H=K_d
  ({\bf n}{\bf S})^2$ which causes size effects for impurities with integer
  spin. Here we show that the dephasing and the energy relaxation are
  influenced by the surface anisotropy in very different
  ways for integer spin having a singlet ground state. That must result also
  in strong size effects and may resolve the puzzle between the
  concentrations estimated from the two kind of experiments.
\end{abstract}

\pacs{73.23.-b,72.15.Qm,71.70.Ej}

\maketitle 

In the present Letter, to our knowledge, is the first time a mechanism is
presented which can resolve the seriously puzzling observation that in some
cases the influence of magnetic impurities on electron dephasing and energy
relaxation have drastically different strengths. The surprising difference in
the estimated impurity concentration raised the doubt about the role and even
the presence of magnetic impurities \cite{GGA1}. That mechanism
is based on surface magnetic anisotropy \cite{UZ1} which is resulting in a
strong size dependence.

The size dependence of the Kondo effect \cite{BG,Roth} was
discovered more than ten years ago and since than it has been carefully
studied experimentally \cite{G,Ha,JG,Se,JG52,JG52yes}. That cannot be
attributed to the size of the Kondo screening cloud reduced by the size of the
sample as only the energy separation of the metallic electron levels are
relevant. That problem was resolved by the suggestion that the
magnetic impurities in the metallic host with strong spin-orbit interaction
experience a surface anisotropy \cite{UZ1}. 

In mesoscopic metallic systems the electron dephasing and energy relaxation
are the central issues in understanding their transport properties
\cite{Birge:1}. The interest has been intensified by the debate over the
saturation of dephasing at low temperature \cite{Mohanty}. The dephasing is
determined e.g. from measurements of magnetoresistance and Aharonov-Bohm rings
in magnetic field \cite{Mohanty,Birge,Natelson,Schopfer} while the energy
relaxation from transport in short wires is found by determining the
nonequilibrium electron energy distribution as shown by the Saclay group
\cite{Pothier}. In many cases the deviations from the expectations of the
theory of Altshuler, Aronov and coworkers \cite{Alt} were attributed to the
presence of magnetic impurities either implanted or contained by the starting
material as contamination. In addition to the energy relaxation due to
electron-electron, electron-phonon interaction the magnetic impurity mediated
electron-electron interaction can play an essential role, which is supported
also by new experiments in magnetic field \cite{Anthore}. It has been known
since a long time \cite{SZ} that such interaction is singular in the energy
transfer $E$ and recently Kaminski and Glazman \cite{KG} called the attention
to similar $1/E^2$ singularity in the electron-electron scattering rate
phenomenologically suggested by the Saclay group. Using that mechanism the
electron transport is determined by the Boltzmann equation and compared with
the experimentally determined electron distributions and the impurity
concentrations were adjusted \cite{th1,th2,th3,th4,GGA1}.  In some cases the
estimated magnetic impurity concentrations using that method are much larger
than those determined from the dephasing rate, even by two orders of
magnitude. In AuPd samples with large spin-orbit interaction the size
dependence of dephasing was also observed \cite{Natelson} where for smaller
size the electron-electron interaction dominates while for larger samples
there are additional scattering mechanism resulting in saturation. That trend
is just the opposite what could be expected in case of additional scattering
centers at the surface. In the following the possible role of surface
anisotropy in these phenomena is discussed.

\begin{figure}
\includegraphics[height=2.8cm]{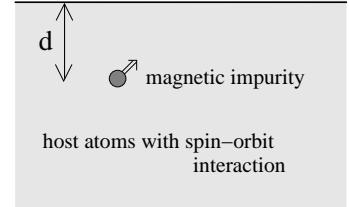}
\caption{\label{fig1} The magnetic impurity at distance $d$ from the surface.} 
\end{figure}

\noindent {\it Surface anisotropy:}\\
For plane-like surfaces (see Fig.~\ref{fig1}) the spin-orbit-induced surface
anisotropy has the
form \cite{UZ1} 
\begin{equation}
  \label{eq:anisham}
  H=K_d ({\bf n}{\bf S})^2
\end{equation}
where ${\bf n}$ is the normal direction of the experienced surface element and
${\bf S}$ is the spin of the impurity. The anisotropy constant $K_d>0$ is
inversely proportional to the distance $d$ measured from the surface
thus it has the form $K_d=\alpha\frac{1}{d}$. 
For thin films using the assumption that the to surfaces of the film
contribute additively, $K_d=\alpha (\frac{1}{d}+\frac{1}{t-d})$, $\alpha$ was
obtained from fitting the Kondo resistivity of Au(Fe) and Cu(Fe) films 
as $\alpha=247$\AA K \cite{UZ2}, of magnetoresistance of Au(Fe)
films as $\alpha=42$\AA K \cite{BZ}, and from multilayer experiments on Au(Fe)
films as $\alpha=60$\AA K \cite{JG}. The parameter $\alpha$ depends also on
the disorder on the surface and in the bulk.  

According to the anisotropy there are different splitting schemes for integer
and half-integer spins. For integer spins (e.g. Fe, Cr $S=2$) the ground 
state is a  
singlet, whereas for half-integer spins (e.g. Mn $S=5/2$) 
it is a Kramers doublet. 
Thus for integer spins the anisotropy causes size effects e.g. in
Kondo resistivity \cite{BG},
magnetoresistance \cite{G}, thermopower \cite{Ha},
impurity spin magnetization \cite{Se}, but for half-integer spins
not \cite{Roth,JG52,UZ3}. 
It is demonstrated that there is a crucial difference between the
cases of integer and half-integer spin. That difference can be less
pronounced for e.g. $S=5/2$ as in the spin glass region pairs or
clusters can be formed which could have also integer spins showing size
dependence \cite{JG52yes}.

\noindent {\it Dephasing:}\\
As the experiments are carried out at low temperature e.g. $T\sim 40$mK, 
thus in
thermal equilibrium most of the higher levels cannot contribute to any
dynamics for impurities with large enough anisotropy, $K_d > kT$. In case of
integer spins the impurity is frozen in a singlet ground state which cannot
lead to dephasing, in contrary to the half-integer spin case where the lowest
states form a doublet and we do see dephasing. Samples with Fe implantation or
contamination must show a strong size dependence in contrary to Mn.

\noindent {\it Energy relaxation:}\\
The nonequilibrium distribution function of a metallic wire with length $L$
and bias $U$ in the diffusive limit is determined by the Boltzmann equation
\begin{eqnarray}
    \frac{\partial f(\ep,x)}{\partial t}&-&\frac{1}{\tau_D}
      \frac{\partial^2 f(\ep,x)}{\partial^2 x} + I_{\mathrm
        coll.}(\{f\})=0\nonumber\\
    I_{\mathrm coll.}(\{f\})&=&\int dE
   \bigl \{f(\ep) [1-f(\ep-E)] W(\ep, E)\nonumber\\
    &-&[1-f(\ep)] f(\ep-E) W(\ep-E,-E) \bigr \}
\end{eqnarray}
where $W(\ep, E)$ is the scattering rate, $\tau_D=\frac{L^2}{D}$ is the
diffusion constant, $f$ is assumed not depending on the spin, and $x$ denotes
the position in the wire in the units of $L$. 
Starting with the solution without inelastic scattering mechanism
\begin{equation}
  \label{eq:freesol}
  f^{(0)}(\ep,x)=(1-x) n_F (\ep-\frac{e U}{2})+x n_F (\ep+\frac{e U}{2})
\end{equation}
and taking into account inelastic scattering in $W$, the Boltzmann equation
can be solved self-consistently at least numerically. 

Here we examine the effect of the surface anisotropy on the energy
relaxation.  Similar to the case of finite magnetic field \cite{GGA1}
the first order processes contribute also to the scattering rate and
the spin occupation numbers $p_M$s depend also on the voltage $U$.
Calculating them from the first order processes we solved the
Boltzmann equation self-consistently using the following collision
integral
\begin{eqnarray}\label{Icoll}
  I^{(2)}_{\mathrm coll.}(\{f\})=&&\hspace{-1.5em}\int dE\int d\ep'
    K^{S}_{MM'} (E,\ep,\ep', K_d) \bigl \{p_{M} f(\ep) f(\ep') \nonumber \\
&&\hspace{-4em}\times [1-f(\ep-E)] [1-f(\ep'+E+K_d M^2
    -K_d M'^2)]\nonumber\\
   &&\hspace{-4em} -p_{M'} [1-f(\ep)] [1-f(\ep')] f(\ep-E) \nonumber \\
 &&\hspace{-4em}\times f(\ep'+E+K_d M^2 - K_d M'^2)\bigr \}
\end{eqnarray}
where the kernel $K^S_{MM'}$ describes electron-electron interaction mediated
by Kondo impurities with surface anisotropy. 

For simplicity we considered the $S=1$ case when the $x$-dependent
$p_M$s are determined from the first order processes by
\begin{equation}
  \label{eq:pms}
  \frac{p_0}{p_1}=\frac{\int d\ep f(\ep,x)
  (1-f(\ep+K_d,x))}{\int d\ep f(\ep,x) (1-f(\ep-K_d,x))}
 \end{equation}
and $2 p_1 + p_0=1$.

\begin{figure}
\centerline{\includegraphics[height=1.5cm]{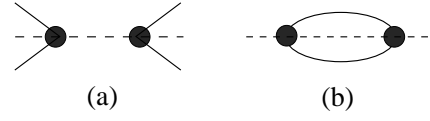}}
\caption{\label{fig2} The diagrams used for calculating (a) the kernel and (b)
  the Korringa lifetime of the impurity spin. The
  solid lines denote the conduction electrons, the dotted lines the impurity
  spin, and the blob is the Kondo coupling.} 
\end{figure}

The kernel in Eq.~(\ref{Icoll}) can be calculated from the diagram
Fig.~\ref{fig2} (a) in the Kondo model 
with anisotropy \cite{UZ2}
\begin{eqnarray}
H&=&\sum\limits_{k,\sigma}\varepsilon_k \,a_{k\sigma}^\dagger
a_{k\sigma}+K_d ({\bf n}{\bf S})^2\nonumber \\
&+& \sum\limits_{\scriptstyle k,k^\prime,\sigma,\sigma^\prime \atop
\scriptstyle M,M^\prime}
J_{MM'} {\bf S}_{MM'}\,(a_{k\sigma}^\dagger \bsigma_{\sigma\sigma'}
a_{k'\sigma'}),
\label{HKondo}
\end{eqnarray}
where $a_{k\sigma}^\dagger$ ($a_{k\sigma}$) creates (annihilates)
a conduction electron with momentum $k$, spin $\sigma$ and energy
$\varepsilon_k$ measured from the Fermi level, 
$\bsigma$ stands for the Pauli matrices and
$J_{MM'}$'s are the Kondo couplings.
The dependence of the interaction kernel $K^{S}_{MM'}$ on the energy transfer
$E$ for $\tau_K=\infty$ is $(E^2)^{-1}$, 
$(E\pm K_d)^{-2}$,
$(E+K_d)^{-1}\cdot(E-K_d)^{-1}$ in different terms, respectively.

For sake of simplicity we used an appropriate constant value ${\tilde J}$
instead of the renormalized Kondo couplings depending on $M,M'$. The influence
of such an approximation was examined in a preceding self-consistent
calculation without surface anisotropy \cite{UJZ}. There the renormalized
coupling was calculated as the solution of the leading logarithmic scaling
equation assuming similar resummation as in equilibrium and smeared by the
spin spectral function with finite Korringa lifetime
$\rho_{s}(\ep)=\frac{1}{\pi}\;\frac{\frac{\hbar}{2\tau_K}}
{\ep^2+\frac{\hbar^2}{4\tau_K^2}}$. The validity of the logarithmic
approximation was always checked by plotting the actual Kondo coupling.  From
the numerical calculations we can conclude \cite{UJZ} that to get the same
results it is a good approximation to replace the renormalized coupling in the
kernel by an appropriately chosen constant value. Furthermore, the smearing of
the renormalized coupling has very small effect \cite{UJZ} on the results for
the parameters consistent with the experimental situation and our results were
in complete agreement with Ref.\cite{GGA1}.

As the weak dependence on the Korringa lifetime \cite{Korringa} $\tau_K$ of
the impurity spin we used the value for $K_d=0$ calculating it from the
diagram Fig.~\ref{fig2} (b) as
\begin{equation}
  \label{eq:tauK}
\frac{\hbar}{2\tau_K (x)}=2\pi (\rho_0 {\tilde J})^2 S (S+1)
\int d\ep (1-f(\ep,x)) f(\ep,x).
\end{equation}
where $\rho_0$ is the conduction electron density of states
for one spin direction.

At each step of the iteration solving the Boltzmann equation self-consistently,
both the spin occupation numbers and the Korringa lifetime were calculated 
from the actual $f$.

\begin{figure}[htb]
\centerline{\includegraphics[height=5.5cm]{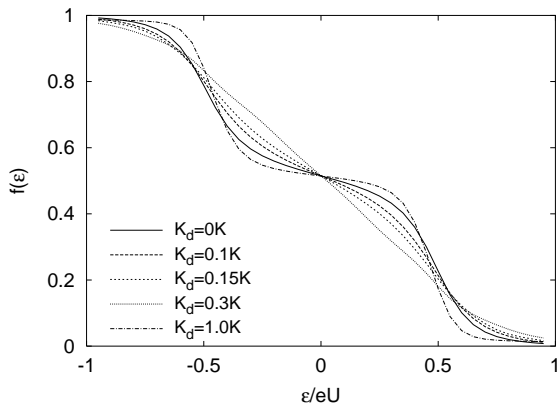}}
\caption{\label{fig3} The calculated distribution function at $x=0.485$
  for different strength of the anisotropy constant $K_d$.  The other
  parameters are $U=0.1$mV, $c=8$ppm, $\rho_0 {\tilde J}=0.11$, and
  $\tau_D=2.8$ns.}
\end{figure}

The dependence of the distribution function on the strength of the anisotropy
constant $K_d$ is illustrated in Fig.~\ref{fig3}. Increasing $K_d$ first the
energy transfer is getting larger but for larger $K_d$ the ground state is
frozen in, similar to the magnetic field dependence discussed in
Ref.~\cite{GGA1}. We can conclude 
that the contribution of magnetic impurities is enhanced
or unchanged in case of finite anisotropy $K_d<e U$.
For $K_d\sim 0.1-0.2$K which is a
good estimation for the strength of the anisotropy for the wires with width of
$\sim 45$nm and thickness of $\sim 85-110$nm used in the experiments, the
energy relaxation is only slightly affected by the anisotropy.

\begin{figure}[htb]
\centerline{\includegraphics[height=5.5cm]{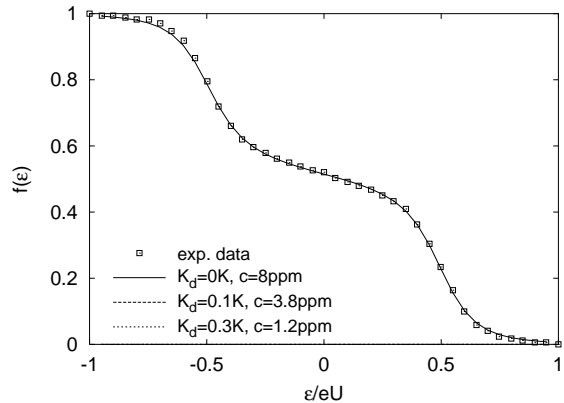}}
\caption{\label{fig4} Fit on the experimental data of Cu wires
  at $x=0.485$ \cite{Anthore} by the calculated distribution function
  for different $K_d$ and $c$ pairs.  The other parameters are
  $U=0.1$mV, $\rho_0 {\tilde J}=0.11$ and $\tau_D=2.8$ns.}
\end{figure}

The goal of the present Letter is not to make optimal fitting of the
experimental curves and determine the value of $K_d$ which must have a
broad distribution itself. We demonstrate, however, that the
experimental curves can be fitted by using combinations of different
values of the concentration $c$ and of $K_d$, and the larger the $c$ is
the smaller the necessary $K_d$. As a demonstration we
compare our results to the experimental data on Cu wires at $x=0.485$
\cite{Anthore} in Fig.~\ref{fig4} as for Cu wires the impurities may
be CuO on the surface having $S=1$ spin \cite{Haesen}. The other fixed
parameters are $U=0.1$mV, $\rho_0 {\tilde J}=0.11$ and $\tau_D=2.8$ns,
and similarly good fits are obtained for $U=0.3$mV as well. The
parameters are somewhat different, which is not surprising as the
distribution for $K_d$ is not taken into account. It is important to
note, that in some cases the origin of the magnetic impurities is not
known, therefore the Kondo temperature corresponding to $\rho_0
{\tilde J}$ in our simple approximation is also a fit parameter.

The half-integer case must be very similar to the case without surface
anisotropy because of the degeneracy, and only the spin dependent prefactors
are different.
 
The two-level system (TLS) may result in somewhat similar behavior \cite{TLS}.
If $T<<\Delta$, where $\Delta$ is the splitting, the dephasing is blocked. In
the nonequilibrium case with applied voltage $U$ ($\Delta < eU$) the spin
dynamics reenters and could lead to dephasing \cite{O} similarly to the
anisotropy case.  Similarly, the energy relaxation becomes also possible but to
get $\frac{1}{E^2}$ singularity at least two non-commuting couplings are
needed \cite{VZ}, thus interaction describing electron screening and electron
induced transition between the levels are required \cite{B}. In this case the
splitting must be small $\Delta < eU$, but the coupling can be weak enough to
be outside the Kondo region. That may result in weak, magnetic field
independent contribution, what is suggested by the experiments \cite{Bi}.

In summary, the surface anisotropy for integer spins is suggested to
reduce drastically the dephasing rate,
while the energy relaxation is much less influenced. In the first case
for low temperature and thermal equilibrium the spin dynamics and
therefore the dephasing are frozen out while in the out-of equilibrium
metallic wire experiments that can reenter. That suggests a pronounced
size dependence and very different concentration for the dynamically
active impurities in the dephasing and the out-of-equilibrium wire
experiments. In the case of half-integer spin having a Kramers doublet
as the lowest state instead of a singlet these cannot be expected.
Further careful experiments for the size dependence and implanted
impurities are required.

\begin{acknowledgments}
We are grateful to N. O Birge, H. Grabert, J. Kroha, H. Pothier, and
G. Zar\'and for useful discussions.
This work was supported by Hungarian grants
OTKA F043465, T046303, T034243, TS040878, T038162, 
and grant No. RTN2-2001-00440.
\end{acknowledgments}

\end{document}